\newcommand{\TP}{\emph{P-T} profile}

\documentclass{iaus}
\usepackage{graphicx}

\title[Atmospheric composition and structure of HD209458b] 
{Atmospheric composition and structure of HD209458b}

\author[J.-M. D\'esert$^1$ et al.]
{J.-M. D\'esert$^1$, A. Vidal-Madjar$^1$, A. Lecavelier des
Etangs$^1$, D.Sing$^1$, D. Ehrenreich$^2$, G. H\'ebrard$^1$ \and
R. Ferlet$^1$}

\affiliation{$^1$Institut d'Astrophysique de Paris, CNRS
(UMR~7095) \\ Universit\'e Pierre \& Marie Curie; 98 bis, boulevard Arago 75014 Paris, France \\ email: {\tt desert@iap.fr} \\[\affilskip]
$^2$Laboratoire d'Astrophysique, Observatoire de Grenoble, UJF, CNRS; \\
BP 53, F-38041 GRENOBLE Cedex 9 (France)}

\pubyear{2008}
\volume{253}  
\pagerange{119--126}
 \jname{Transiting planets} \editors{F.
Pont, eds.}

\begin{document}

\maketitle

\begin{abstract}
Transiting planets like HD209458b offer a unique opportunity to
scrutinize their atmospheric composition and structure. Transit
spectroscopy probes the transition region between the day and
night sides, called limb. We present a re-analysis of existing
HST/STIS transmission spectra of HD209458bs atmosphere. From
these observations we: Identify H2 Rayleigh scattering, derive the
absolute Sodium abundance and quantify its depletion in the upper
atmosphere, extract a stratospheric T-P profile with a
temperature inversion and explain broad band absorptions with the
presence of TiO and VO molecules in the atmosphere of this planet.

\end{abstract}

\firstsection 
\section{Introduction}

Because of the wavelength-dependent opacities of absorbing
species, measurement of relative changes in eclipse depth as a
function of wavelength during primary transit has the potential to
reveal the presence (or absence) of specific chemical species
(Seager et al.\ 2000a, Hubbard et al.\ 2001, Brown et al.\ 2001).

In the case of HD~209458b's atmosphere, the transmission spectroscopy
method led  to the detection of sodium (Charbonneau et~al. 2002,
Sing et~al. 2008b). In the UV, absorptions of several percents for
H\,{\sc i} Lyman-$\alpha$, O\,{\sc i} and C\,{\sc ii} have been
measured in the hydrodynamically escaping upper atmosphere
(Vidal-Madjar et al.\ 2003, 2004, 2008, D\'esert et al.\ 2004,
Lecavelier\ 2007, Ehrenreich et al. 2007).

We use public archived {\it STIS} spectra obtained during
planetary transit at two spectral resolutions (low and medium).
Both datasets are combined to extend the measurements over the
entire optical regime to quantify possible absorbers appearing in
the transmission spectrum (Sing et al. 2008a).

In this poster we present the identification of Rayleigh
scattering by H$_2$ molecules (Lecavelier et~al. 2008b), the
depletion of NaI as well as the extraction of a \TP\ with
inversion (Sing et~al. 2008b). Finally, we derived the upper
limits for the TiO/VO molecular abundances in the atmosphere of
the planet (D\'esert et al. 2008).

\section{Observed features in the spectrum and interpretation}

The \emph{HST-STIS} G750L, and G430L low resolution grating
observations (2003) of HD~209458b analyzed here are also detailed in
Knutson et al.\ (2007a), Barman (2007), Ballester et al.\ (2007)
and Sing et al. (2008a). For both the G750L, and G430L gratings,
two visits were done for each grating, of five consecutive orbits
each. Together they cover the combined range 2900-10300 \AA, with
some overlap around 5300 \AA\ with a resolving power $R=500$.

\begin{figure}[b]
\begin{center}
 \includegraphics[angle=90,width=2.4in]{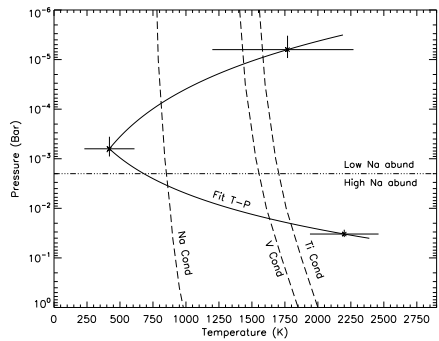}
 \caption{
The atmospheric Temperature-Pressure profile (\TP) with error
bars. The dashed lines correspond to the condensation curves for
Titanium, Vanadium and Sodium. The three points of this T-P
profile are derived from the fit of the observed absorption depth
curve. The hot point at a pressure of 0.05 Bar is imposed by the
Rayleigh scattering (See Lecavelier et al. 2008).}
   \label{fig_TP}
\end{center}
\end{figure}

\begin{figure}[b]
\begin{center}
 \includegraphics[width=3.in]{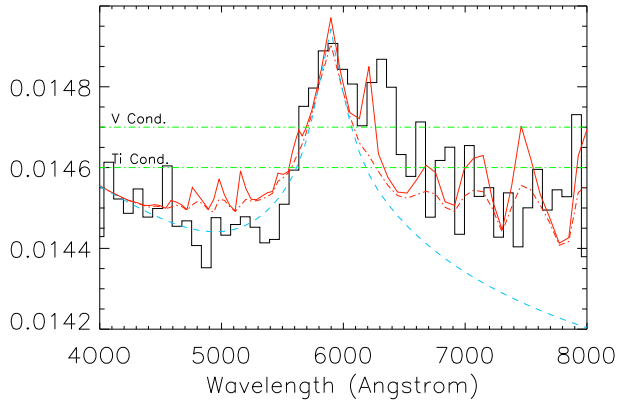}
 \caption{
The low resolution STIS measurements of the planetary transit
absorption depth (AD) corrected from limb-darkenning effects and
binned by 60 pixels (histogram). The dashed line corresponds to
the best fit model assuming Rayleigh scattering and sodium
absorption with the physical T-P profile plotted in
Fig.~\ref{fig_TP}. Overplotted in continuous line, is the same
model, with best fit models assuming Rayleigh scattering, NaI, TiO
and VO absorptions. The dot-dash line correspond to the model with
a cold-trap. The two altitudes of condensation for the TiO and VO
molecules are plotted horizontally.}
   \label{fig1}
\end{center}
\end{figure}
Lecavelier Des Etangs et al. (2008a) show that the observed
Absorption Depth (AD) is well approximated by:

\begin{equation}
AD_{\lambda}=AD_0(1+\frac{2H}{R_P}\ln\frac{\sigma_{\lambda}}{\sigma_{\lambda_0}})
\label{ADbin}
\end{equation}

\noindent where $AD_0$ is AD at $\lambda=\lambda_0$ and H the
scale height. Thus the observed mean AD over a wavelength range of
a given spectral element, is proportional to the temperature and
to the logarithm of the cross section.

We use the absorption depth (AD) curve as a function of the
wavelength from 4000 \AA\ to 8000 \AA\ obtained by Sing et al.\
2008a. The spectrum at wavelength below 8000 \AA\ is considered
here where the absorption due to water molecules is negligible
(Barman et al.\ 2007). This AD spectrum is composed of three
remarkable features (see Fig.~\ref{fig1}):

\

\begin{enumerate}

\item \textit{Rayleigh scattering}. The feature in NUV, at
wavelength $\lambda$ $\le$ 5000 \AA, was first reported by
Ballester et al.\ 2007 and explained by the absorption of a hot
hydrogen layer within the atmosphere. Here we propose an
alternative explanation invoking the Rayleigh scattering by H$_2$
molecules (Lecavelier et al.\ 2008b).

\

\item \textit{Na depletion and atmospheric structure}. Within the
same datasets, we find that the Na spectral line profile is
characterized by a wide absorption with a sharp transition to a
narrow absorption profile at higher altitudes values (Sing et al.\
2008b). This sharp transition is interpreted by condensation or
ionization which deplete Na atoms in the upper atmosphere. Using a
global fit to these data, from 3000 \AA\ to 6200 \AA, we determine
the average pressure-temperature profile ($P-T$), see
Fig.~\ref{fig_TP}, at the planetary terminator (Sing et al.\
2008b).

\

\item \textit{TiO/VO}. A third broad band absorption feature in
the range 6200 \AA\ $-$ 8000 \AA\ appears on the absorption depth
spectrum and is still unexplained. Although no typical TiO and VO
spectral signatures have been identified unambiguously in the
observed spectrum, we suggest that the opacities of those
molecules are the best candidates to explain the remaining
continuous broad band absorption observed in this wavelength
domain (D\'esert et al. submitted). Using the \TP\ from Sing et
al.\ 2008b, we derived upper limits for the TiO and VO abundances.
We found that the abundance of TiO should be around $10^{-4}$ to
$10^{-3}$ solar, and  the abundance VO around $10^{-3}$ to
$10^{-2}$ solar (D\'esert et al. submitted).

\end{enumerate}

\section{The Irradiated Atmospheres of HD209458b}

The temperature inversion leads to high temperature at both low
and high pressure. This temperature bifurcation was very well
predicted by atmospheric models of strongly irradiated planets (
Hubeny et al.\ 2003). In the lower part of the atmosphere
($\sim$30 mbar), the temperature is found to be in the range $1900
-2400$ K (Lecavelier et~al. 2008b), which corresponds to the M/L/T
brown dwarf regime, as expected for a hot Jupiter such as HD~209458b
(Kirkpatrick et al.\ 2005).

However, depending on the effective temperature, a large number of
diatomic and polyatomic molecules are predicted to be present
according to various models of brown dwarf and hot Jupiter
(Burrows et al.\ 1999, Lodders et al.\ 1999, Allard et al.\ 2001,
Lodders et al.\ 2002, Hubeny et al.\ 2003). Among those molecules
and at a temperature above 1800 K, titanium oxyde (TiO) and
vanadium oxyde (VO) in gas phase equilibrium are most probably
present with a high abundance in strongly irradiated planet
atmospheres (Seager et al.\ 1998, Hubeny et al.\ 2003, Fortney et
al.\ 2007). Furthermore, the low albedo measurements of HD~209458b (Rowe
et al.\ 2006) rule out most of the absorbants, but TiO and VO.
Recently, it has been found that a theoretical fit to the HD~209458b
near infrared observation at secondary eclipse requires that the
dayside atmosphere of HD 209458b have a thermal inversion and a
stratosphere (Knutson et al. 2008, Burrows et al.\ 2007c, Burrows
et al. 2008). Hubeny et al.\ (2003) and Fortney et al.\ (2007)
highlight the importance of gaseous TiO and VO opacity in their
model of highly irradiated close-in giant planets. The last
authors define two classes of irradiated atmospheres. Those which
are warm enough to have a strong opacity due to TiO and VO gases
(``pM Class'' planets), and those that are cooler (``pL Class''
planets) dominated by Na\,{\sc i} and K\,{\sc i}. Our possible
detection of TiO/VO and Na\,{\sc i} in the hot atmosphere of HD~209458b
confirm that this planet is located in the transition region
between the two classes defined by these authors.

Further observation are necessary to better characterize the two
different type planets using TiO/VO. More planets can be likely
studied using transit observations which allowed discoveries,
detection, and characterization of extrasolar objects such as
planets and comets (Lecavelier des Etangs et al.\ 1995, 1999a,
2005).


\begin{thebibliography}{}


\bibitem[{{Allard} {et~al.} 2001}]{Allard01}
Allard, F., Hauschildt, P.~H., Alexander, D.~R., Tamanai, A., \&
Schweitzer, A.\ 2001, \textit{ApJ}, 556, 357

\bibitem[{{Ballester} {et~al.} (2007)}]{Ballester07}
{Ballester}, G.~E., {Sing}, D.~K., \& {Herbert}, F. 2007,
\textit{Nature}, 445, 511

\bibitem[{{Barman} 2007}]{Barman07}
{Barman}, T. 2007, \textit{ApJ} (Letters), 661, L191

\bibitem[{{Brown} 2001}]{Brown01}
{Brown}, T.~M. 2001, \textit{ApJ}, 553, 1006

\bibitem[{{Burrows} \& {Sharp} 1999}]{Burrows99}
{Burrows}, A. \& {Sharp}, C.~M. 1999, \textit{ApJ}, 512, 843

\bibitem[{{Burrows} {et~al.} 2007c}]{Burrows07c}
Burrows, A., Hubeny, I., Budaj, J., Knutson, H.~A., \&
Charbonneau, D.\ 2007, \textit{ApJ} (Letters), 668, L171

\bibitem[{{Burrows} {et~al.} 2008}]{Burrows08}
Burrows, A., Budaj, J., \& Hubeny, I.\ 2008, \textit{ApJ}, 678,
1436

\bibitem[{{Charbonneau} {et~al.} (2002)}]{Charbonneau02}
Charbonneau, D., Brown, T.~M., Noyes, R.~W., \& Gilliland, R.~L.\
2002, \textit{ApJ}, 568, 377

\bibitem[{{Charbonneau} {et~al.} 2005}]{Charbonneau05}
Charbonneau, D., Allen, L.~E., Megeath, S.~T., et al.\ 2005,
\textit{ApJ}, 626, 523

\bibitem[2004]{Desert et al.2004}D{\'e}sert, J.-M., Vidal-Madjar, A., Lecavelier
Des {\'E}tangs, A., et al.\ 2004, Extrasolar Planets: Today and
Tomorrow, 321, 205

\bibitem[2008]{Desert2008}D\'esert, J.-M., Lecavelier
des Etangs, A., Vidal-Madjar, A., Sing, D.~K., et al.\ 2008,
\textit{A\&A}, submitted

\bibitem[{{Ehrenreich} {et~al.} 2008}]{Ehrenreich08}
Ehrenreich, D., Lecavelier Des Etangs, A., H{\'e}brard, G., et
al.\ 2008, \textit{A\&A}, 483, 933

\bibitem[{{Fortney} {et~al.} 2003}]{Fortney03}Fortney, J.~J., Sudarsky, D., Hubeny, I.,
et al.\ 2003, \textit{ApJ}, 589, 615

\bibitem[{{Fortney} {et~al.} 2006b}]{Fortney06}Fortney, J.~J., Cooper, C.~S., Showman,
A.~P., Marley, M.~S., \& Freedman, R.~S.\ 2006, \textit{ApJ}, 652,
746

\bibitem[{{Fortney} {et~al.} 2007}]{Fortney07}
Fortney, J.~J., Lodders, K., Marley, M.~S., \& Freedman, R.~S.\
2008, \textit{ApJ}, 678, 1419

\bibitem[{{Hubbard} {et~al.} 2001}]{Hubbard01}
Hubbard, W.~B., Fortney, J.~J., Lunine, J.~I., et al.\ 2001,
\textit{ApJ}, 560, 413

\bibitem[{{Hubeny} {et~al.} 2003}]{Hubeny03}
Hubeny, I., Burrows, A., \& Sudarsky, D.\ 2003, \textit{ApJ}, 594,
1011

\bibitem[{{Knutson} {et~al.} (2007a)}]{Knutson07a}
Knutson, H.~A., Charbonneau, D., Noyes, R.~W., Brown, T.~M., \&
Gilliland, R.~L.\ 2007, \textit{ApJ}, 655, 564

\bibitem[{{Knutson} {et~al.} 2008}]{Knutson08}
Knutson, H.~A., Charbonneau, D., Allen, L.~E., Burrows, A., \&
Megeath, S.~T.\ 2008, \textit{ApJ}, 673, 526

\bibitem[{{Kirkpatrick} 2005}]{Kirkpatrick05}
{Kirkpatrick}, J.~D. 2005, \textit{ARAA}, 43, 195

\bibitem[1995]{Lecavelier Des Etangs et al.1995}Lecavelier Des Etangs, A.,
Deleuil, M., Vidal-Madjar, A., et al.\ 1995, \textit{A\&A}, 299,
557

\bibitem[1999a]{Lecavelier Des Etangs et al.1999a}Lecavelier Des Etangs, A.,
Vidal-Madjar, A., \& Ferlet, R.\ 1999, \textit{A\&A}, 343, 916

\bibitem[2007]{Lecavelier Des Etangs2007}Lecavelier Des Etangs, A.\ 2007,
\textit{A\&A}, 461, 1185

\bibitem[2008]{Lecavelier Des Etangs et al.2008}Lecavelier Des Etangs, A.,
Vidal-Madjar, A., D{\'e}sert, J.-M., \& Sing, D.\ 2008,
\textit{A\&A}, 485, 865

\bibitem[{{Lodders} 1999}]{Lodders99}
{Lodders}, K. 1999, \textit{ApJ}, 519, 793

\bibitem[{{Lodders} 2002}]{Lodders02}
{Lodders}, K. 2002, \textit{ApJ}, 577, 974

\bibitem[2000]{Nitschelm et al.2000}Nitschelm, C., Lecavelier des Etangs,
A., Vidal-Madjar, A., et al.\ 2000, \textit{A\&AS}, 145, 275

\bibitem[{{Rowe} {et~al.} 2006}]{Rowe06}
Rowe, J.~F., Matthews, J.~M., Seager, S., et al.\ 2006,
\textit{ApJ}, 646, 1241

\bibitem[1998]{Seager et al.1998}Seager, S.,
\& Sasselov, D.~D.\ 1998, \textit{ApJ}l, 502, L157

\bibitem[{{Seager} \& {Sasselov} 2000}]{Seager2000a}
{Seager}, S. \& {Sasselov}, D.~D. 2000, \textit{ApJ}, 537, 916

\bibitem[{{Sing} {et~al.} (2008a)}]{Sing08a}
Sing, D.~K., Vidal-Madjar, A., Desert, J.~-., Lecavelier des
Etangs, A., \& Ballester, G.\ 2008, ArXiv e-prints, 802,
arXiv:0802.3864

\bibitem[{{Sing} {et~al.} (2008b)}]{Sing08b}
Sing, D.~K., Vidal-Madjar, A., Lecavelier des Etangs, A., et al.\
2008, ArXiv e-prints, 803, arXiv:0803.1054

\bibitem[{{Vidal-Madjar} {et~al.} 2003}]{Madjar03}Vidal-Madjar, A., Lecavelier des
Etangs, A., D{\'e}sert, J.-M., et al.\ 2003, \textit{Nature}, 422,
143

\bibitem[{{Vidal-Madjar} {et~al.} 2004}]{Madjar04}Vidal-Madjar, A., D{\'e}sert, J.-M.,
Lecavelier des Etangs, A., et al.\ 2004, \textit{ApJ} (Letters),
604, L69

\bibitem[{{Vidal-Madjar} {et~al.} 2008}]{Madjar08}Vidal-Madjar, A., Lecavelier des
Etangs, A., D{\'e}sert, J.-M., et al.\ 2008, \textit{ApJ}
(Letters), 676, L57


\end{thebibliography}
\end{document}